\documentclass[reprint,amsmath,amssymb,aps]{revtex4-2}
\usepackage{graphicx}
\usepackage{dcolumn}
\usepackage{bm}

\usepackage{xcolor}

\begin{document}

\preprint{APS/123-QED}

\title{Self-propulsion via slipping: frictional swimming in multi-legged locomotors}

\author{Baxi Chong$^{1,2}$, Juntao He$^{3}$, Shengkai Li$^{2}$, Eva Erickson$^{2}$, Kelimar Diaz$^{1,2}$, Tianyu Wang$^{3}$, Daniel Soto$^{3}$, Daniel I. Goldman$^{1,2,3}$}

\affiliation{
{$^1$}{Interdisciplinary Graduate Program in Quantitative Biosciences}\\
{$^2$}{School of Physics}\\
{$^3$}{Institute for Robotics and Intelligent Machines}\\
Georgia Institute of Technology, Atlanta, GA, USA
}%

\begin{abstract}


Locomotion is typically studied either in continuous media where bodies and legs experience forces generated by the flowing medium, or on solid substrates dominated by friction. In the former, centralized coordination is believed to facilitate appropriate slipping through the medium for propulsion. In the latter, slip is often assumed minimal and thus avoided via decentralized controls. We discover in laboratory experiments that terrestrial locomotion of a meter scale multi-segmented/legged robophysical model resembles undulatory fluid swimming. Experiments varying waves of limb stepping and body bending reveal how these parameters result in effective terrestrial locomotion despite seemingly ineffective isotropic frictional contacts. Dissipation dominates over inertial effects in this macroscopic-scaled regime, resulting in essentially geometric locomotion akin to microscopic-scale swimming. Theoretical analysis demonstrates that the high-dimensional multi-segmented/legged dynamics can be simplified to a centralized low-dimensional model, which reveals an effective Resistive Force Theory with an acquired viscous drag anisotropy. We extend our low-dimensional, geometric analysis to illustrate how body undulation can aid performance in non-flat obstacle-rich terrains and also use the scheme to quantitatively model how body undulation affects performance of biological centipede locomotion (the desert centipede \textit{S. polymorpha}) moving at relatively high speeds ($\sim 0.5$ body lengths/sec). Our results could facilitate control of multilegged robots in complex terradynamic scenarios.

\end{abstract}

\maketitle



Locomotion by body undulation is often observed in locomotors continuously immersed in an environment (such as fluids or granular media)~\cite{gray1946mechanism,jayne1986kinematics,socha2002gliding,hatton2010generating,marvi2012friction,kern2006simulations,jung2010caenorhabditis}. 
During such self-propulsion, body elements continuously experience forces set by the physics of the medium and the instantaneous orientations and velocities of body elements. 
An approach for analyzing such locomotion, which integrates thrust and drag forces over the body, was introduced in the early to mid 20th century and goes by Resistive Force Theory (RFT). This method has successfully modeled organisms in highly damped hydrodynamic and granular terradynamic environments, like microorganisms and sand-swimmers etc~\cite{maladen2009undulatory,schiebel2019mechanical,schiebel2020mitigating,li2013terradynamics}. 
RFT works at its core because of a so-called "drag anisotropy" as elements translate through continuous media. For example, long thin systems like spermatoza undulating in fluids can be thought of as a superposition of slender rods, which differ in reaction forces in the perpendicular and parallel directions~\cite{becker2003self,avron2004optimal,tam2007optimal,fu2007theory,crowdy2011two}.

In contrast, on solid terradynamic environments like flat ground, a key feature of the locomotion dynamics is that the anatomical elements (e.g., legs or body segments) are no longer in constant contact with the environment. Rather, these elements can make and break contact~\cite{hu2009mechanics,jing2013optimization}. Interaction models in such situations are typically assumed to be rate-independent isotropic Coulomb friction. In such situations, control algorithms~\cite{buchli2009compliant,kajita2004biped} were developed to minimize slip\footnote{(Sliding between the the substrate and the leg. Note that we only consider slipping occurring at the tip (foot) of the leg which interacts with the substrate.} during leg contact; similarly, such active slip avoidance is also observed in biological systems~\cite{clark2011slipping,marigold2003role}. Slipping is actively avoided partially because Coulomb friction introduces a step-function between the velocity-friction relationship, which can cause unstable oscillations~\cite{elmer1997nonlinear}. Furthermore, if not properly controlled, slipping can reduce the energetic efficiency~\cite{zhou2014energy}.

\begin{figure}[h!]
\includegraphics[width=1\linewidth]{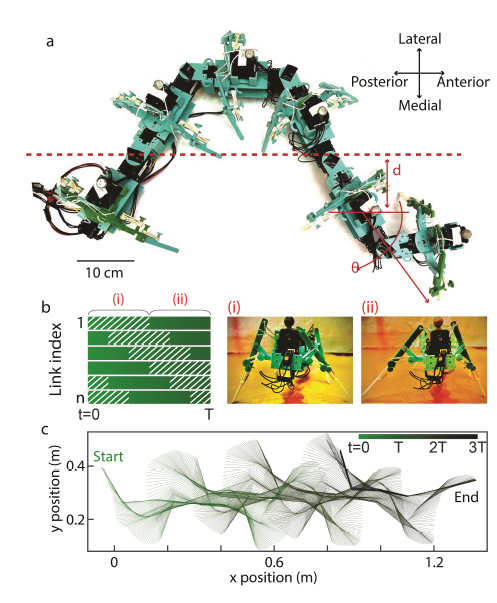}
\caption{\label{fig:intro} \textbf{Swimming in terrestrial environments} (a) Oblique view of the (\textit{top}) desert centipede, \textit{S. polymorpha}, and (\textit{bottom}) an overhead view of the robophysical device. (b) The patterns of lifting and landing of contralateral feet. Each row represents the contact states of $i-th$ link. Shadow region represents right foot in stance phase, open region presents the left foot in stance phase. (right) Front view of the robophysical device lifting (i) left and (ii) right feet of the first module  (c) Trajectory of backbone during terrestrial swimming ($\Theta_{body}=\pi/3$, $\Theta_{leg}=0$, $n=6$) colored by time. (d) (\textit{Left}) Displacement and (\textit{right}) velocity profile of terrestrial swimming. We compare the experimental data with dynamical simulation (brown curve) and quasi-static simulation (blue curve) for body-dominated terrestrial swimming. After the transient development, both experiments' dynamical simulation converge to a limit cycle similar to quasi-static simulation.}
\end{figure}

\begin{figure}[h!]
\includegraphics[width=1\linewidth]{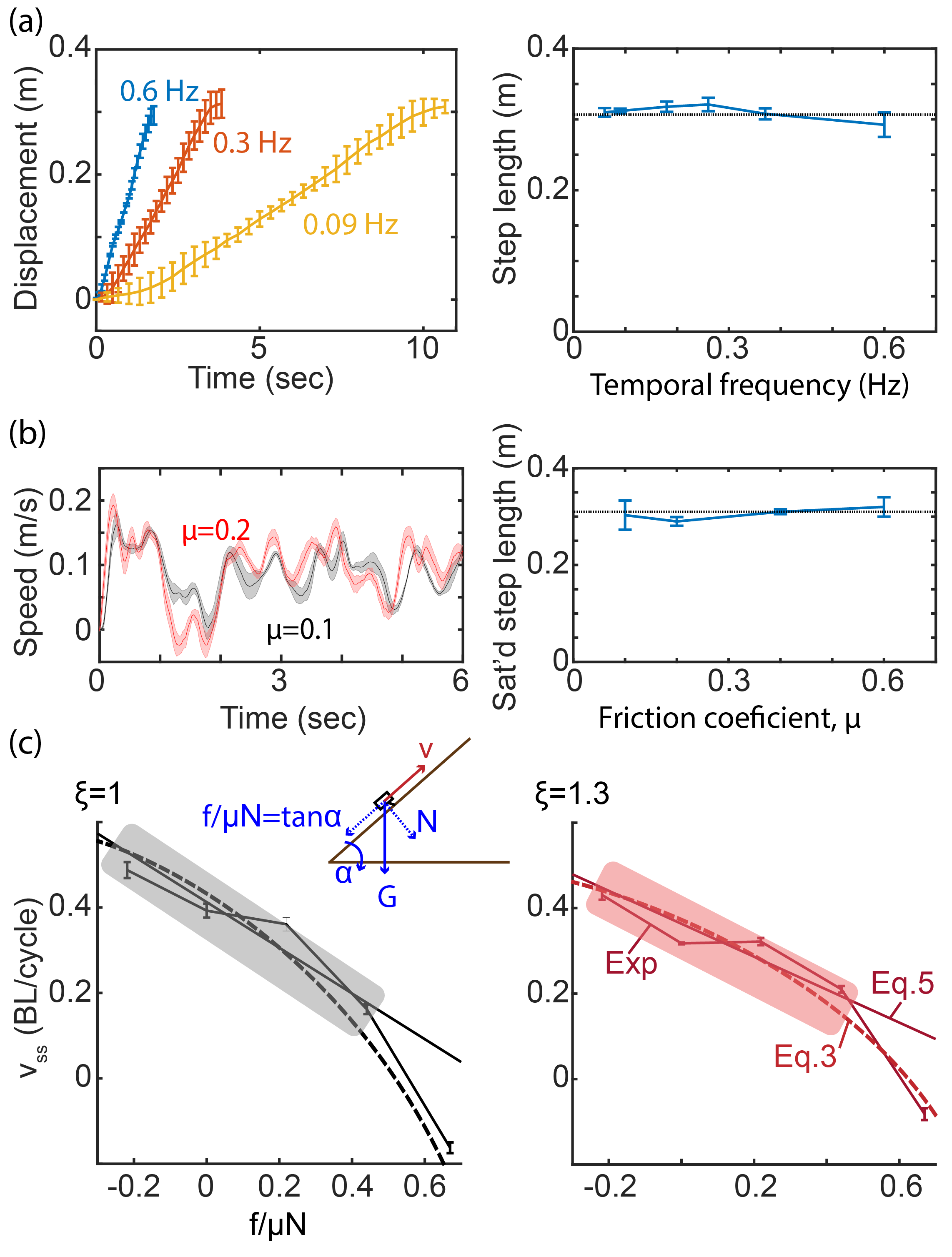}
\caption{\label{fig:ada} \textbf{The geometric nature of terrestrial swimming} (a) Robot ($n=6$) implementing the same gait ($\Theta_{body}=\pi/3$, $\Theta_{leg}=0$) under different temporal frequencies. (\textit{Left}) The development of CoG displacement as a function of time under different temporal frequencies. (\textit{Right}) The step length is stable over a range of temporal frequencies. Dashed lines represent prediction from quasi-static simulation. (b) Robot implementing the same gait under different substrate (different friction coefficients, $\mu$). (\textit{Left}) The development of CoG velocity as a function of time. Despite the initial high-magnitude oscillation, robots on low-friction surfaces converged to quasi-static velocity profiles after one gait cycle. (\textit{Right}) The saturated step length is stable over a range of friction coefficients. (c) Experimental verification of force-velocity relationship ($\Theta_{body}=\pi/3$, $\Theta_{leg}=0$, $n=8$). We test the relationship between the whole body drag force and the velocity by measuring the speed of robots on slopes. We compare experimental results (curves with error bar) and theoretical predictions from Eq.~\ref{eq:3} (dashed curves) and~\ref{eq:5} (solid curves) for two spatial frequencies, $\xi=1$ and $\xi = 1.3$. In both experiments, we observe that there exists a linear relationship between force and velocity near equilibrium.}
\end{figure}

\begin{figure}[h!]
\includegraphics[width=1\linewidth]{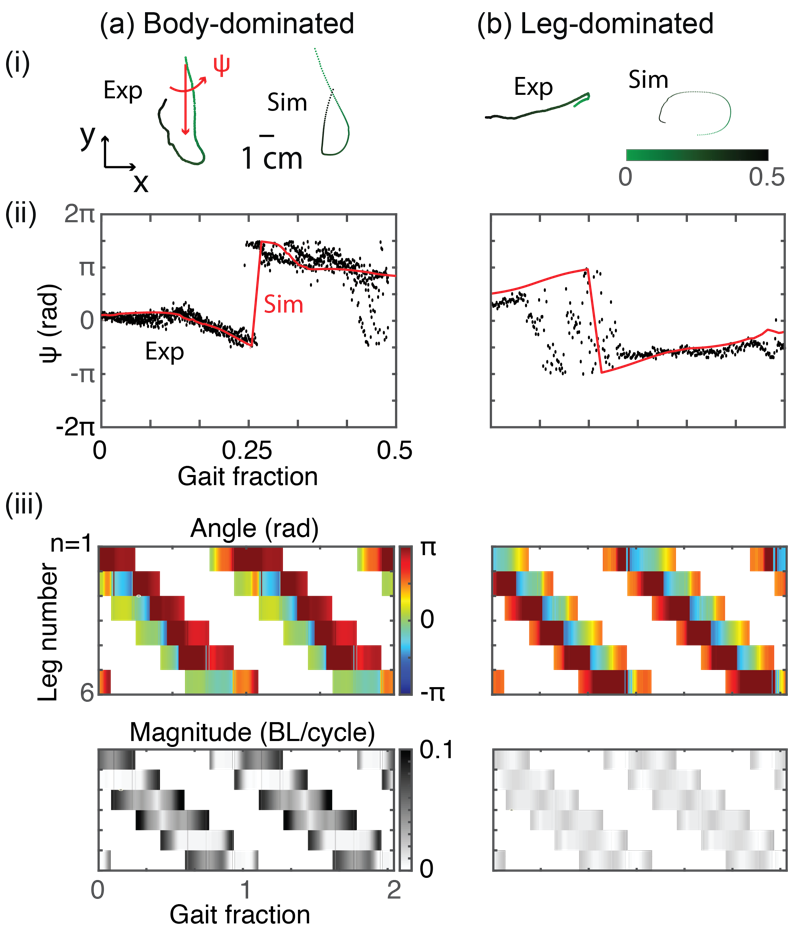}
\caption{\label{fig:slipang}  \textbf{Direction of foot slipping in body-dominated and leg dominated gaits} (i) Typical trajectories of a foot (second foot from the left) of a robophysical device ($n=6$) during stance phase for (a) body-dominated ($\Theta_{body}=\pi/3$, $\Theta_{leg}=0$) and (b) leg-dominated ($\Theta_{body}=0$, $\Theta_{leg}=\pi/6$) terrestrial swimming. $x$-axis is the direction of motion. We quantify the slipping of a foot by its direction ($\Psi$, unit: rad) and magnitude (unit: BL/cycle). (ii) The simulation prediction (red curves) and experimental measured (black dots) time series of slipping angles. (iii) The slipping profiles from simulation.  We illustrate (\textit{top}) the slipping direction profile and (\textit{bottom}) the slipping magnitude profile. }
\end{figure}

\begin{figure*}[t]
\includegraphics[width=1\linewidth]{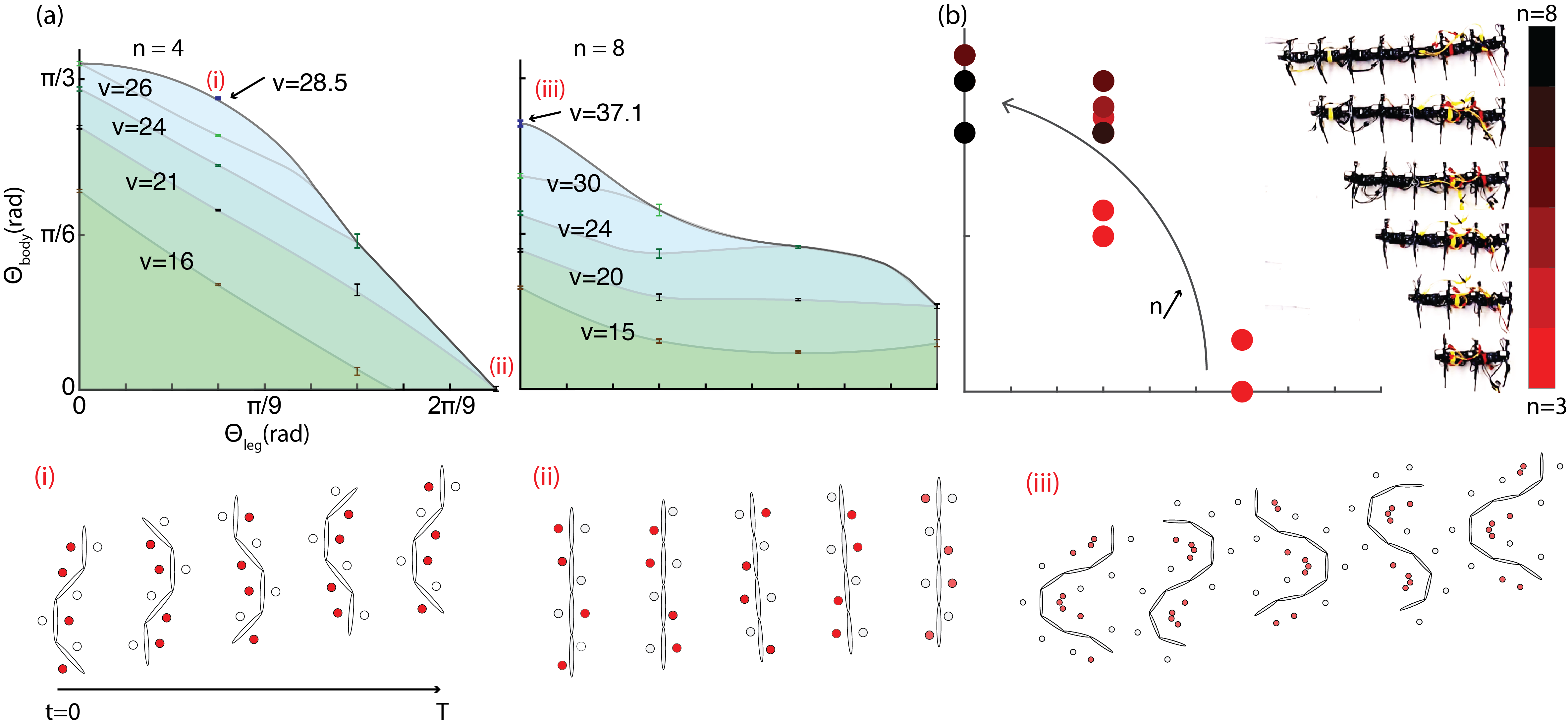}
\caption{\label{fig:phasediagram} \textbf{Performance space of multi-legged terrestrial locomotion.}  We characterize the terrestrial locomotion using a performance space consisting of the amplitudes of body undulation and leg movement.  (a) The heat map of velocity ($v$, unit: cm/cycle) over the performance space for robophysical model with (\textit{left}) 4, (\textit{right}) 8 pairs of legs. $n$ is the number of leg pairs. Note that optimal locomotion (the highest velocity) occurs at "hybrid" region when $n=4$, and at body-dominated region when $n=8$. (\textit{bottom}) Snapshots of body configurations over a cycles for (i) $n=4$, $\Theta_{body}=60^\circ$, $\Theta_{leg}=15^\circ$, (ii) $n=4$, $\Theta_{body}=0^\circ$, $\Theta_{leg}=45^\circ$, and  (iii) $N=8$, $\Theta_{body}=60^\circ$, $\Theta_{leg}=0^\circ$. (b) The transition of optimal terrestrial locomotion from leg-dominated to body-dominated as the number of leg pairs increases.}
\end{figure*}

\begin{figure}[h!]
\includegraphics[width=1\linewidth]{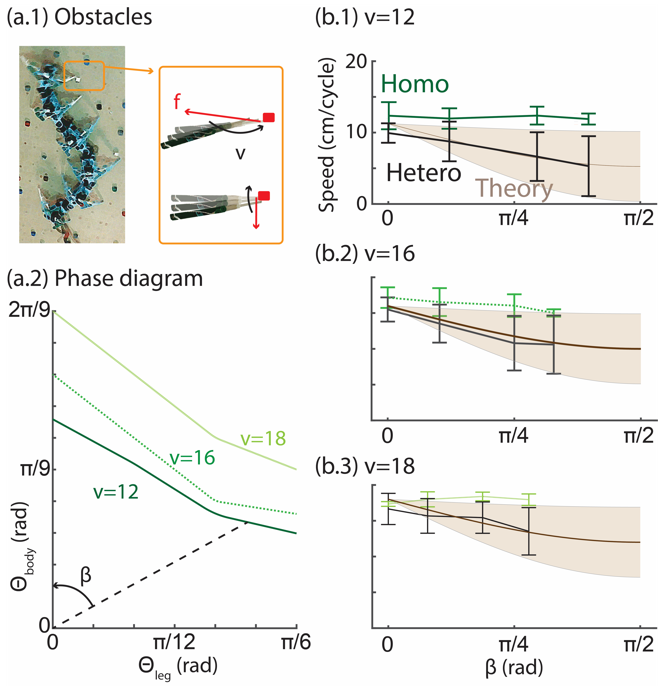}
\caption{\label{fig:oal} \textbf{Advantage of body-dominated terrestrial swimming in obstacle-rich environments} (a.1) A snapshot of robot ($n=6$) moving on obstacle-rich environments ($\rho=0.06$). Cartoon illustration of interaction between robot and obstacles subject to different slipping directions (\textit{top}: typical body-dominated; \textit{bottom}: typical leg-dominated). Red blocks represents obstacles, legs from darker color to lighter color represents progression of time. (a.2) We choose three isoheight lines on the velocity heat-map over the performance space: $v=12$~cm/cycle, $v=16$~cm/cycle, and $v=18$~cm/cycle. We quantify the degree of body and leg use by $\beta = \tan^{-1}({\Theta_{leg}}/{\Theta_{body}})$.  (b) Comparison of locomotion performance in homogeneous (curves with error bar in green colors), heterogeneous environments (curves with error bar in black color), and theoretical predictions from Eq. 6 (curves and areas in light brown color). From top to bottom the flat line descended from isoheight lines with $v=18$cm/cycle, $v=18$cm/cycle, and $v=18$cm/cycle.}
\end{figure}

While the bulk of prior work on terrestrial locomotors~\cite{alexander1982locomotion} focused on systems with two or four legs, many biological, and increasingly robotic devices, possess multiple sets of limbs (e.g., cockroaches have six and centipedes can have up to 40 legs). In contrast to the few-legged systems in which an assumption of no-slip contact is often feasible~\cite{westervelt2018feedback}, for systems with more than four legs, there is a high possibility that some slip has to occur during locomotion~\cite{zhao2020multi} because of kinematic constraint violations, (e.g., the BigAnt~\cite{zhao2020multi}). We hypothesize that, instead of avoiding slipping, multi-legged locomotors can actively coordinate slips for effective propulsion and we can analyze such behavior using a method similar to RFT on continuous media, and establish a unifying model for locomotion on both terrestrial and continuous media. The challenges lie in the nonlinearity and the isotropy of Coulomb friction in terrestrial environments in contrast to the linear, anisotropic viscous friction.

Here, we investigate the slipping and thrust-generation mechanism in multi-legged locomotors where both body undulation and leg retraction contribute to slipping and self-propulsion. Using a centipede-like robophysical model~\cite{chong2022coordinating}, we show that the steady-state terrestrial locomotion has a property of geometric locomotion (the effect of inertia is negligible) even when operated at high speed on a low-friction substrate. We use RFT to study slipping in multi-legged systems and propose a new principle of acquired drag anisotropy. Specifically, by periodic lifting and landing of body appendages, the nonlinear and isotropic Coulomb friction experienced on each leg can be recast into a velocity-dependent whole-body drag, similar to that of organisms at a low Reynolds number, which we refer to as "terrestrial swimming". In an effort to unify our proposed slip-driven mechanism and the conventional minimal-slipping mechanism, we establish a performance space of terrestrial swimming, and discuss the relative advantage (i.e., higher speed and more robust to obstacle-rich environments) of body-dominated (slip-driven) over leg-dominated (minimal slipping) terrestrial swimming by robophysical experiments. Finally, we use our scheme to analyze the locomotion of a biological multi-legged system (centipedes) and reveal its slip-driven terrestrial swimming behavior. Similar to our predictions of robophysical experiments, we observe a smooth gait transition from leg-dominated to body-dominated locomotion as speed increases.

\section*{Body undulation}

We adapt a robophysical modeling approach~\cite{aydin2019physics} to systematically study terrestrial locomotion. Specifically, we construct a multi-legged robot consisting of repeated modules. Each module contains one pair of legs and one body connection. All combined, each module has three degrees of freedom (DoF): the shoulder lifting joint that controls the contact states of contralateral legs, the shoulder retraction joint that controls the fore/aft positions of leg movements, and the body bending joint that controls the lateral body undulation~\cite{ozkan2020systematic}. 
The synchronization of these three DoF is coupled using the extended Hildebrand framework (\cite{chong2022general} and SI), which prescribe a leg stepping wave and a body undulation wave, both propagating from head to tail. The amplitude of body undulation, $\Theta_{body}$, the amplitude of leg movement, $\Theta_{leg}$, and the spatial wave number $\xi$, can then uniquely prescribe the gait of the multi-legged robot. Note that unless otherwise mentioned, we set $\xi=1$ throughout the paper.

As discussed in prior work, body undulation can play an important role in multi-legged systems~\cite{chong2022general,chong2021coordination}. In Fig.~\ref{fig:intro}c, we show the midline trajectory during undulatory locomotion of the multi-legged robot ($\Theta_{body}=\pi/3$, $\Theta_{leg}=0$, $n=6$, $n$ is the number of modules). Depending on the relative magnitude of $\Theta_{body}$ and $\Theta_{leg}$, we soft-classify the performance space into (1) body-dominated ($\Theta_{body}\gg \Theta_{leg}$), (2) hybrid ($\Theta_{body}\sim \Theta_{leg}$), and (3) leg-dominated ($\Theta_{body}\ll \Theta_{leg}$). While the body parts are lifted off the ground, we note that the undulatory body trajectory is similar to limbless slithering motion (commonly observed in snakes and nematodes locomoting on continuous media such as sand~\cite{astley2020surprising} and viscous fluid~\cite{berri2009forward}). 

To quantitatively investigate terrestrial swimming, we track the trajectory of the center of geometry (CoG) of the robophysical model. In Fig.~\ref{fig:intro}d left panel, we illustrate the displacement and speed profile. Interestingly, we observe that after the transient response ($t<2s$) upon the initiation of gait, the trajectory of velocity converges to a limit cycle. To better understand the initial transient response and the limit cycle, we establish a dynamic model (see SI) and a quasi-static model~\cite{chong2022general} (Fig.~\ref{fig:intro}d). While the dynamic model under-predicted\footnote{We posited that it is the static friction that leads to the discrepancy between the empirically measured and model predicted transient response.} the magnitude of the transient response upon the initiation of gait, the predicted velocity from this dynamic model also converges to a limit cycle. The average of the speed over limit cycle is almost identical in experiments, dynamic model prediction and quasi-static model prediction, indicating that inertial effects are not significant in the dynamic system in terrestrial swimming.

To quantify the effect of inertia, we test the locomotion performance of the multi-legged system ($\Theta_{body}=\pi/3$, $\Theta_{leg}=0$, $n=6$) under different temporal frequencies (Fig.~\ref{fig:ada}a). We show that, despite the changes in absolute speed, ranging from $\approx15$ cm/s to 1.5 cm/s, the step length (body length traveled per cycle) is almost constant. Furthermore, we test the locomotion performance of the multi-legged systems ($\Theta_{body}=\pi/3$, $\Theta_{leg}=0$, $m=6$) on different surfaces ranging from coarsely fabricated wood ($\mu\sim0.6$) to coated smooth surfaces ($\mu\sim0.1$). Across all surfaces, swimming motion converges to the steady-state equilibrium velocity within one gait cycle (Fig.~\ref{fig:ada}b). 

We posit that such convergence to steady-state equilibrium velocity can be a result of an emergent  friction-velocity negative feedback. To explore this force-velocity relationship, we test the locomotion performance on slopes. Specifically, by varying the slope tilting angle $\alpha$, we can measure the relationship between the external force $\tan{\alpha}$ (normalized by nominal friction $\mu N$) and the step length. We test two undulatory gaits with different spatial wave numbers ($\Theta_{body}=\pi/3$, $\Theta_{leg}=0$, $m=6$, $\xi=\{1, 1.3\}$). In both cases, we observe a locally negative linear relationship (Fig.~\ref{fig:ada}c) between external force and step length. The emergence of such negative linear relationship not only explains the convergence, but also raises an intuiting concept: effective viscous friction emerges from terrestrial swimming  with Coulomb friction. In the next sections, we will further analyze and model such emergent negative linearity.

\section*{Effective viscous friction}

\subsection*{Slipping analysis}

Similar to locomotion at a low Reynolds number, we consider terrestrial swimming as an undulatory swimming system with assistance from the periodic leg lifting and landing. As documented in prior work on locomotion at a low Reynolds number~\cite{koens2016rotation,jing2013optimization}, the drag anisotropy of slender rods (higher reaction forces in the perpendicular than in the parallel direction) is the critical physical property enabling swimming in viscous flows. In terrestrial environments where drag force is typically assumed isotropic Coulomb friction, the direct implementation of undulatory motion would be ineffective~\cite{alben2019efficient}. 

In Coulomb friction, the direction of ground reaction forces should be opposite to the direction of slipping. Therefore, it is crucial to investigate the direction of slipping. Unlike other legged systems with fewer legs, there is significant slipping during undulatory locomotion of the multi-legged robot. We predict from the quasi-static model that in body-dominated gaits, the slipping is predominantly in the lateral direction. 
We verify this prediction by tracking the trajectory of the tip of a foot (second foot from the left) and empirically measuring the direction of slipping (Fig.~\ref{fig:intro}) when operating a body-dominated gait ($\Theta_{body}=\pi/3$, $\Theta_{leg}=0$, $n=6$). 
We quantify the direction of slipping by measuring the slipping angle $\Psi$, defined as the angle between the direction of slipping and the medial direction. We compare the experimentally measured and simulation predicted time series of slipping angles in Fig.~\ref{fig:slipang}a.ii, and both suggest that the direction of foot slipping is almost always perpendicular to the direction of motion ($\Psi=0$ or $\pi$). Finally, we show the slipping angle profile from numerical simulation in Fig.~\ref{fig:slipang}a.iii. We notice that for almost all feet, the slipping angle is distributed around either $0$ or $\pi$, both suggesting lateral/medial slipping.

\subsection*{Kinematic model}

With the knowledge of lateral/medial dominated slipping, we develop a theoretical model to illustrate how periodic leg lifting and landing can acquire drag anisotropy similar to locomotors in viscous flow.

As documented in prior work on undulatory locomotion, each body segment experiences oscillation in the lateral and rotational direction with an offset of $\pi/2$~\cite{maladen2011mechanical,rieser2019geometric}. Specifically, $d$, the distance from the body to the central body axis can be expressed as: $d(\tau)=d_m \sin{\tau}$, where $d_m$ is the magnitude of lateral oscillation and $\tau\in [0\ 2\pi )$ is the gait phase; $\theta$, the angle between the body orientation and the direction of motion can be expressed as $\theta(\tau) = \theta_m \cos{\tau}$, where $\theta_m$ is the magnitude of rotational oscillation (Fig.~\ref{fig:intro}a). $\theta_m$ and $d_m$ are determined by the amplitudes, ($\Theta_{leg}$ and $\Theta_{body}$) and the spatial wave number ($\xi$) of body undulation. From geometry, we know that $d_m = n\Theta_{body}/(2\pi\xi)^2$, and $\theta_m = \Theta_{leg}+\tan^{-1}\big(n\Theta_{body}/(2\pi\xi)\big)$. 

To simplify our analysis, we assume that the center of geometry (CoG) of the robot has a constant forward velocity, $v$. The velocity of a foot on the right hand side of body\footnote{For simplicity, we only discussed the right feet. The analysis of left feet will be symmetric to our analysis.} can then be expressed as a joint effect of CoG movement and the lateral/rotational oscillation:

\begin{align}
    v_x(\tau) = & \ \dot{d}(\tau) + l\dot{\theta}(\tau)\sin{\big(\theta(\tau)\big)} \nonumber \\
    v_y(\tau,v) = & \ v +  l\dot{\theta}(\tau)\cos{\big(\theta(\tau)\big)}
\end{align}

\noindent where $v_x$ and $v_y$ are velocity components in the lateral and anterior directions respectively; $l$ is the leg length. Friction should have the opposite direction to the direction of foot slipping. Thus, the projection of the instantaneous frictional force to the anterior direction is:

\begin{align}
    f_y(\tau,v) & = -\mu N \sin{\big(\tan^{-1}{(\frac{v_y(\tau,v)}{v_x(\tau)})}\big)} 
\end{align}

\noindent where $\mu N$ is the magnitude of friction determined by the normal force $N$ and the friction coefficient $\mu$. Assuming that each contralateral foot is in contact with the substrate for half of a period (e.g., $s_1<\tau<s_1+\pi$), we can calculate the average friction over the stance phase:

\begin{align}\label{eq:3}
     \bar{f}(v)  = \int_{s_1}^{s_1+\pi} -\mu N \sin{\big(\tan^{-1}{(\frac{v_y(\tau,v)}{v_x(\tau)})}\big)} d\tau
\end{align}

We can calculate the steady-state CoG velocity, $v_{ss}$, by assuming the force in equilibrium ($\bar{f}(v_{ss})=0$). By the force balance, $\bar{f}(v_{ss})=0$, we establish an implicit function $v_{ss} = v_{ss}(s_1)$. Furthermore, we take a variational approach to find the optimal stance period $[s_1,\ s_1+\pi]$ to maximize $v_{ss}$ (i.e., $dv_{ss}/ds_1 =0$). The sufficient condition for $s_1$ (to optimize $v_{ss}$) is then:

\begin{equation}
    \sin{\big(\tan^{-1}{(\frac{v_y(s_1+\pi)}{v_x(s_1+\pi)})\big)} = \sin{\big(\tan^{-1}{(\frac{v_y(s_1    )}{v_x(s_1    )})}\big)}},
\end{equation}

Solving Eq. 4 yields two optima: $s_1 = 0$, $s_1 = \pi$. They correspond to maximal $v_{ss}$ (highest forward speed) and minimal $v_{ss}$ (highest backward speed) respectively. In other words, by properly controlling the sequence of lifting and landing, we can effectively acquire drag anisotropy in either direction and therefore enable swimming along (direct wave~\cite{manton1952evolution}) and against (retrograde wave~\cite{manton1952evolution}) the direction of wave propagation. Interestingly, $s_1 = 0$ also optimizes body-leg coordination as reported in \cite{chong2022general} where the body undulation is considered to assist leg retraction. In this paper, we only consider retrograde-wave terrestrial swimming. Thus we set $s_1 = 0$ unless otherwise stated.

Since slipping is primarily in the lateral direction, we assume $v_x \gg v_y$. We can therefore calculate the changes in friction in response to disturbance to steady state velocity ($v=v_{ss}+\delta_v$):

\begin{align}
    f_y(\tau,v_{ss}+\delta_v) &  = -\mu N \sin{\big(\tan^{-1}{(\frac{\delta_v +  v_y(\tau,v_{ss})}{v_x(\tau)})}\big)} \nonumber\\
    [\because v_x \gg v_y]\ \ \ \ & \approx f_y(\tau,v_{ss})  -\mu N \sin{\big(\tan^{-1}{(\frac{v_{ss}      }{v_x(\tau)})}\big)}\frac{\delta_v}{v_{ss}},     \nonumber                
\end{align}

\noindent Integrating over the stance period, we can obtain the changes of the average friction:
\begin{align}\label{eq:5}
     \underbrace{\bar{f}(v_{ss}+\delta_v)}_{\bar{f}_d(\delta_v)}  & = \underbrace{\bar{f}(v_{ss})}_{0}- \delta_v \underbrace{\int_{0}^{\pi} \frac{\mu N}{v_{ss}}\sin{\big(\tan^{-1}{(\frac{v_{ss}      }{v_x(\tau)})}\big)} d\tau}_{\gamma_0} \nonumber \\
     \bar{f}_d(\delta_v) & = -\gamma_0\ \delta_v
\end{align}

The effective linear force-velocity relationship allows us to analyze the terrestrial swimming similar to that in viscous fluid. 
Despite being counter-intuitive with Coulomb friction, Eq. 5 predicts that this equilibrium is asymptotically stable. Note that our analysis is invariant to the choice of foot. Eq. 3 and 5 can thus be generalized to the overall multi-legged system by a scaling factor of $n$ (the number of leg pairs). To verify our analysis, we compare predictions from Eq. 3 and Eq. 5 to the experimental measurement in Fig.~\ref{fig:ada}c, and we observe good agreement between theory and experiments, especially locally near equilibrium.



\section*{Performance space}

As discussed earlier, both body undulation and leg retraction can contribute to generate thrust in multi-legged systems. To systematically explore the coordination and balance of body and leg, we introduce a performance space (Fig.~\ref{fig:phasediagram}) where the axes are $\Theta_{leg}$ and $\Theta_{body}$. Note that competition exists between high $\Theta_{body}$ and high $\Theta_{leg}$ since it will lead to self-collision among legs which can break the robot. In previous discussions, we focused on the body-dominated terrestrial swimming regime of the performance space. The conventional leg-dominated counter-part experiments ($\Theta_{body}=\pi/3$, $\Theta_{leg}=0$, $n=6$) are provided in Fig.~\ref{fig:slipang}b. Note that slipping in conventional leg-dominated gaits is significantly lower than those in body-dominated terrestrial swimming.

To systematically explore the competition and coordination between body undulation and leg pair reduction, we experimentally tested the locomotion performance of different points on the performance space. Fig.~\ref{fig:phasediagram}a shows a heat-map of speed over performance space for robots with 4 (Fig.~\ref{fig:phasediagram}a. \textit{left}) and 8 (Fig.~\ref{fig:phasediagram}a. \textit{right}) pairs of legs. Immediately, we notice that for systems with different leg pairs, the optima resides in different regimes. For robots with 4 pairs of legs, a hybrid mode of body undulation and leg retraction can lead to the highest speed. On the other hand, for robots with 8 pairs of legs, pure body-dominated terrestrial swimming ($\Theta_{leg}=0)$ can led to the highest speed. This is also evident in the gradient of iso-height contours. To further quantify the transition, we identify\footnote{We numerically determine the optima as over 90 percentile of the step length among all gaits.} the optima, $[\Theta_{leg},\ \Theta_{body}]$, for robots with from 3 to 8 pairs of legs. We then color the optima $[\Theta_{leg},\ \Theta_{body}]$ by the number of leg pairs $n$. From Fig.~\ref{fig:phasediagram}b, we observe that the optima transition from leg-dominated to body-dominated as the number of leg pairs increases.

\section*{Interaction with obstacles}

In this section, we further explore the relative advantage of body-dominated and leg-dominated gaits in obstacle-rich environments. We posit that the slipping direction plays an important role in the interaction with obstacles; and that body-dominated gaits (with lateral/medial slipping) are more robust over the presence of obstacles as compared to leg-dominated gaits (with anterior/posterior slipping).

In Fig.~\ref{fig:slipang} we compare the direction of slipping for leg-dominated and body-dominated gaits. Specifically, slipping in leg-dominated gaits almost always occurs first in the anterior direction then follows in the posterior direction. This chronological order of slipping can affect the interaction with terrain heterogeneity (obstacles). In other words, the interaction between a leg and an obstacle is more likely to occur during the preceding slipping than the succeeding slipping. The presence of obstacles can interfere with slipping and therefore offer a reaction force opposite to the direction of slipping. Thus, the preceding anterior slipping feet in conventional leg-dominated gaits can be detrimental to locomotion. On the other hand, in body-dominated gaits, feet slip in lateral/medial directions, in which reactions from interactions with obstacles are also in medial/lateral direction and will not affect the locomotion performance in the direction of motion.

To verify this prediction, we construct a heterogeneous environment (low-height obstacles randomly distributed on a flat terrain, see SI) and test the locomotion performance of different gaits in the multi-legged system ($n=6$). We identify three iso-height lines on the performance space such that all points on an iso-height line have the same step length on homogeneous environment. We choose the iso-height lines with $v=12$, $v=16$, and $v=18$ (unit cm/cycle). We quantify the degree of body and leg use by the angle $\beta = \tan^{-1}(\frac{\Theta_{leg}}{\Theta_{body}})$. Interestingly, we notice that gaits with higher $\beta$ have significantly reduced step length in heterogeneous environments than in homogeneous environments. 


To better understand the robustness of gaits in heterogeneous environments, we establish a simple statistical model. To simplify the analysis, we approximate the slipping angle $\Psi$ (Fig.~\ref{fig:slipang}) by $\beta$ such that $\Psi=0$ during body-dominated terrestrial swimming ($\beta=0$) and $\Psi=\pi/2$ during conventional leg-dominated terrestrial swimming ($\beta=\pi/2$). Assuming the reaction force from terrain heterogeneity is constant, $F$, the projection of reaction force in fore-aft direction can be approximated by $F\sin{(\beta)}$. Consider an obstacle-rich environment with obstacle density $\rho$ and a robot with $n$ pairs of legs, then the distribution function of at least one leg interacting with an obstacle is $h(y)=\{1\ \text{if}\ y<n\rho;\ 0\ \text{if}\ y\ \geq n\rho\}$, where $y\sim U(0,1)$, $U$ is the uniform distribution. Thus the distribution function of the projection of reaction force in fore-aft direction is $Fh(y)\sin{\beta}$. From Eq. 5, the distribution function of step length is:

\begin{align}
    v & \sim v_{ss}- \gamma_0^{-1}F\sin{(\beta)}h(y) \nonumber \\
    \bar{v} &= v_{ss}- \gamma_0^{-1}F(1-n\rho)\sin{(\beta)} \nonumber \\
    \text{std}(v) &= \gamma_0^{-1}F\sqrt{n\rho(1-n\rho)}\sin{(\beta)},
\end{align}

\noindent where $\gamma_0$ is the effective drag coefficient from Eq. 5. We observe quantitative agreement between the theoretical prediction and the experiments (Fig.~\ref{fig:oal}).

\section*{Biological centipedes}


\begin{figure}[h!]
\includegraphics[width=1\linewidth]{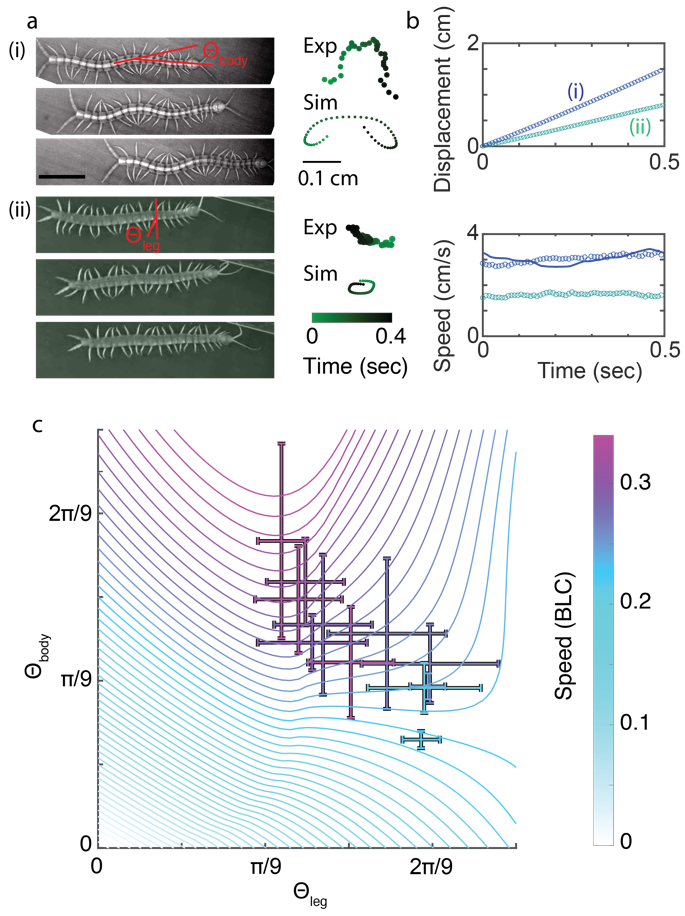}
\caption{\label{fig:animal} \textbf{Analysis of centipede  locomotion} (a) (\textit{left}) Snapshots of a trial of (i) body-dominated and (ii) leg-dominated centipede locomotion. (\textit{right}) The trajectory of foot slipping colored by time. The stance phase spans 0.4 seconds.Trajectory of centipede body during terrestrial swimming colored by time. (b) The (\textit{top}) displacement and (\textit{bottom}) velocity profiles. Measured animal data is presented in cycles and the predictions from quasi-static model is presented in blue curve. (c) (Left) Body undulation and (right) leg movement profile. Dashed black curve labeled the propagation of body wave and leg wave along the body. Axes are identical in all panels. (d) The transition from leg-dominated gaits to swimming dominated gaits as speed increases. In each centipede locomotion trial, we extract amplitude of body undulation and leg retraction, and represented it as a cross (colored by its speed) on performance space.}
\end{figure}


Biological centipedes coordinate their body and leg movement to rapidly traverse different terrestrial environments~\cite{manton1952evolution,anderson1995axial,dutta2019programmable,kuroda2022gait}. However, there has been limited biomechanical analysis on centipede locomotion. Centipedes possess a large number of legs and body segments. Such high dimensionality can make the direct biomechanical modeling difficult. Therefore, one of the challenges in analyzing centipede locomotion lies in the proper dimensionality reduction. Despite recent efforts in dimensionality reduction in artificial multi-legged robots~\cite{chong2022coordinating}, it remains unclear if such dimensionality reduction can be extended to biological centipedes. However, if identified, such low dimensional representation of centipede kinematics will not only simplify the biomechanical analysis (over original high-dimensional motion control) but also pave the way towards the understanding of the possible centralized neuromechanical control. Further, existing work on centipedes typically assumes that there is no foot slipping~\cite{yasui2017decentralized,yasui2019decoding,aoi2016advantage}. Given the importance of slipping as illustrated in our robophysical experiments, we posited that slipping also plays an important role in biological centipede locomotion despite the potential low magnitude. 

Here, we use our low-dimensional slipping model to study rapidly-moving biological centipedes. We predict from our drag anisotropy analysis (solving Eq. 3) that body-dominated gaits should be a faster mode of locomotion as compared to leg-dominated (Fig.~\ref{fig:animal} the underlying heatmap). The model predicts that at high $\Theta_{body}$, low $\Theta_{leg}$, centipedes could maintain high-speed steady-state motion with feet slipping laterally/medially. However, if $\Theta_{leg}$ is increased, it will introduce unwanted anterior/posterior slipping, which can break the symmetry in the steady-state swimming motion and therefore is detrimental to locomotion.

To test this prediction, we study the locomotion performance of a biological centipede (\textit{Scolopendra polymorpha}, 5 individuals, 11 trials in total) and characterize the leg dynamics under different speeds. Specifically, we collect high speed video recordings of centipedes moving on a white board. We noticed that the wave propagation in both body and leg has negligible variance in the maximum amplitude (Fig.~\ref{fig:animal}c), which allows us to present the whole-body kinematics using a low-dimensional performance space. We extract the amplitude of body undulation ($\Theta_{body}$) and leg movements ($\Theta_{leg}$), and the speed (in the unit of body length per cycle) from each trial (presented by colored cross in Fig.~\ref{fig:animal}d). We notice that the emergence of body undulation is accompanied by the decrease in $\Theta_{leg}$. This indicates that in response to high speeds, the behavior of these centipedes is beyond just the emergence of body undulation. Instead, there is a transition of leg-dominated gaits (high $\Theta_{leg}$, low $\Theta_{body}$) to body-dominated gaits (low $\Theta_{leg}$, high $\Theta_{body}$), in agreement with our prediction. 

Further, we use our model to study the kinematics behind centipede locomotion. Similar to our analysis on the robot, we compare a body-dominated gait (Fig.~\ref{fig:animal}a.i) and a leg-dominated gait (Fig.~\ref{fig:animal}a.ii) in biological centipedes. We then investigate the direction of foot slipping for both cases. Interestingly, we observe that slipping is extensive and mostly in the lateral/medial direction for body-dominated gaits. In contrast, slipping is reduced and in the anterior/posterior direction for leg-dominated gaits. Finally, we show the displacement and velocity profile for both body-dominated and leg-dominated gaits in Fig.~\ref{fig:animal}b. Notably, our low-dimensional quasi-static model can give a quantitative prediction of the velocity profile for the relatively high-speed ($\sim0.5$ BL per second) centipede terrestrial swimming, reconciling the low-dimensional centralized control nature of seemingly complicated multi-legged systems.

\section*{Conclusions}

In this paper, we performed systematic tests of multilegged locomotion and discovered that its dynamics were highly damped. We then developed a theory to explain these observations and further used this theory to quantitatively model the locomotion of biological centipedes.

From a physics of locomotion perspective, this paper reveals that a unified framework can capture and explain undulatory swimming in highly damped environments with both intrinsic and acquired drag anisotropy. Drag anisotropy is believed to be the critical principle which enables effective undulatory swimming~\cite{gray1955propulsion,jing2013optimization,koens2016rotation,alben2019efficient,alben2021efficient}. Prior work has typically considered the intrinsic property of an element translating through a flowable medium (e.g., viscous fluid~\cite{koens2016rotation} and granular media~\cite{li2013terradynamics,Maladen:2011es}) or a locomotors' surface structures~\cite{hu2009mechanics,rieser2021functional} as the cause of drag anistropy and therefore effective undulatory swimming~\cite{gray1955propulsion,gray1953undulatory,jayne1988muscular,Maladen:2011es,stephens2008dimensionality}. More recently, studies have demonstrated effective undulatory swimming with no intrinsic drag anisotropy. These works modulated the magnitude of surface traction via either static friction~\cite{alben2019efficient,jing2013optimization} or periodic lifting and landing of body appendages~\cite{chong2022general} to produce undulatory locomotion. Here, we posit that effective undulatory swimming shares the same physical principles between these two scenarios (intrinsic drag anisotropy and friction modulation). Specifically, we demonstrated that terrestrial multi-legged locomotion can be recast as a fluid-like problem with the nonlinearites of the foot-ground interaction leading to acquired drag anisotropy in the environments dominated by isotropic, rate-independent Coulomb friction. Notably, the body-leg coordination to optimize leg retraction converges to the leg use to maximize acquired drag anisotropy. By performing these experiments and developing this framework, we broaden and deepen our understanding of undulatory swimming and allow for comparison and cross-referencing of locomotion in different substrates.

This paper further reveals the geometric nature of terrestrial swimming. Typically observed in highly-damped environments, geometric locomotion has a property that net translation is generated from properly coordinated self-deformation to counter drag forces~\cite{kelly1995geometric,ostrowski1998geometric,shapere1989geometric,hatton2015nonconservativity}. During the last decades, physicists and engineers have developed a powerful geometric mechanics framework to understand biological undulatory swimming behaviors~\cite{rieser2019geometric,astley2020surprising,chong2022coordinating} and design novel swimming gaits for robots~~\cite{shammas2007geometric,hatton2015nonconservativity,hatton2013geometric}. This geometric approach replaces laborious calculation with illustrative diagrams and therefore offers quantitative and qualitative insights into locomotion. However, the application of geometric mechanics was limited to the environments where friction dominated over the inertial forces. Surprisingly, recent work successfully used geometric mechanics to study legged systems~\cite{chong2022general,ozkan2020systematic,zhao2020multi} despite not having a solid theoretical foundation (non-negligible inertia). Here, we showed that terrestrial swimming has a property of effective viscous drag, which guaranteed the convergence to steady-state quasi-static locomotion despite the temporal frequency and friction coefficient. In this way, our framework rationalizes the application of geometric mechanics to multi-legged terrestrial swimming, offering building blocks for future exploration of terrestrial swimming using geometric mechanics. 


From an engineering perspective, our robophysical studies and RFT scheme can aid in the development of multi-segmented legged robots. If properly controlled, robots with different body morphologies and limb numbers could be used in different tasks. For example, legged robots are known for their agility~\cite{hwangbo2019learning,pratt1998intuitive,saranli2001rhex} whereas multi-segmented (e.g., serially connected legless) robots for interactions with obstacles~\cite{transeth2008snake,wang2020omega,wang2022generalized}. Multi-segmented legged robots have the potential to leverage the advantages of both multi-segmented and legged robots~\cite{ozkan2021self}. However, to date, there has been no effective tool to control such devices to the full measure of their potential. With the help of our RFT framework and robophysical systematic experiments, our scheme paves the way towards an alternative for robust and agile locomotion.

Finally, our work can also contribute to a new understanding of the neuromechanics of myriapod biological locomotion. With redundant legs, biological centipedes possess high mobility in diverse environments~\cite{manton1952evolution,yasui2019decoding,anderson1995axial}. Additionally, some centipedes are reported to display body undulation at high speeds~\cite{manton1952evolution,anderson1995axial}. 
Manton hypothesized that the use of body undulation is passive and thus detrimental to locomotion efficiency~\cite{manton1952evolution}. In this way, Manton further posited that centipedes with the capability to resist body undulation at high speeds are evolutionarily advanced~\cite{manton1952evolution}. Over the past century, the mechanism of body undulation remains controversial. Using body-mechanical dynamic modeling, recent work~\cite{aoi2016advantage,aoi2013instability} suggested that the presence of body undulation could be a passive outcome of dynamic instability. On the contrary, electromyography (EMG) data analysis suggests that activity of the axial musculature causes body undulations, indicating an active nature of body undulation~\cite{anderson1995axial}. 
Independent of a passive or active mechanism, the role (e.g., whether beneficial/detrimental) of body undulation during locomotion remains less studied. Here, our observation and analysis indicates that there exists a transition from low-step-length gaits to high-step-length gaits as speed increases, and such transition includes two essential steps: the emergence of body undulation and the reduction in leg amplitude. Thus, our model illustrates the essential role of body undulation to aid locomotion at high speed, offering insights into the neuromechanical and evolution studies of different centipedes.

\section*{Materials and Methods}


\subsection*{Robophysical experiments}

All of the robophysical models were designed in Solidworks and 3D printed (LulzBot TAZ Workhorse) using PLA material. Each module in multi-legged robots has a pair of rigidly connected legs (12 cm in length). There are three servo motors (Robotis Dynamixel AX-12a) in each segment: one controls horizontal body bending and the other two control the fore/aft and up/down motion of the legs. Servo motors are powered with an external power supply (11.2 V) and communicate with PC via a micro controller (Robotis U2D2). During a single experiment, all gait parameters in the gait that the robot execute are constant, and servo motor set points are sent at a fixed frequency (33 Hz for the lowest robot speed, and 200 Hz for the highest speed). In each experiment, we implemented the gaits on robots for at least 5 trials (homogeneous) and 10 trials (heterogeneous), each trial has 3 cycles.

To capture the position and orientation of the robot, we attached a reflective marker on each module of the robot and used an Optitrack motion capture system (four 360 FPS, Prime 17W cameras, and software Motive) to track positions of the markers in the workspace. The tracked data was analyzed in MATLAB.

\subsection*{Gait prescription of multi-legged robots}

We use a binary variable $c$ to represent the contact state of a leg, where $c=1$ represents the stance phase and $c=0$ represents the swing phase. Following~\cite{chong2022general}, the contact pattern of symmetric quadrupedal gaits can be written as

\begin{align}
     c_l(\tau_c,1) &=\begin{cases}
      1, & \text{if}\ \text{mod}(\tau_c,2\pi) < 2\pi D\\
      0, & \text{otherwise}
    \end{cases} \nonumber \\
    c_l(\tau_c,i) &= c_l(\tau_c - 2\pi\frac{\xi}{n}(i-1),1) \nonumber \\
    c_r(\tau_c,i) &= c_l(\tau_c+\pi,i),   
    \label{eq:general_hild}
\end{align}

\noindent where $\xi$ denotes the number of spatial waves on legs, $D$ the duty factor, $c_l(\tau_c,i)$ (and $c_r(\tau_c,i)$) denotes the contact state of $i$-th leg on the left (and the right) at gait phase $\tau_c$, $i\in \{1, ... n\}$ for $2n$-legged systems (See Fig. S1). 

Legs generate self-propulsion by protracting during the stance phase to make contact with the environment, and retracting during the swing phase to break contact. 
That is, the leg moves from the anterior to the posterior end during the stance phase and moves from the posterior to anterior end during the swing phase. 
With this in mind, we use a piece-wise sinusoidal function to prescribe the anterior/posterior excursion angles ($\theta$) for a given contact phase ($\tau_c$) defined earlier,
\begin{align}
        \theta_l(\tau_c,1)  &=\begin{cases}
      \Theta_{leg}\cos{(\frac{\tau_c}{2D})}, & \text{if}\ \text{mod}(\tau_c,2\pi)  < 2\pi D\\
      -\Theta_{leg}\cos{(\frac{\tau_c-2\pi D}{2(1-D)})}, & \text{otherwise},
    \end{cases} \nonumber \\
    \theta_l(\tau_c, i) &= \theta_l(\tau_c - 2\pi\frac{\xi}{n}(i-1), 1) \nonumber \\
    \theta_r(\tau_c, i) &= \theta_l(\tau_c + \pi, i) 
    \label{eq:legmove}
\end{align}

\noindent where $\Theta_{leg}$ is the shoulder angle amplitude, $\theta_l(\tau_c,i)$ and $\theta_r(\tau_c,i)$ denote the leg shoulder angle of $i$-th left and right leg at contact phase $\tau_c$, respectively.  Note that the shoulder angle is maximum ($\theta=\Theta_{leg}$) at the transition from swing to stance phase, and is minimum ($\theta=-\Theta_{leg}$) at the transition from stance to swing phase. Note that we chose $D=0.5$ unless otherwise mentioned.

We then introduce lateral body undulation by propagating a wave along the backbone from head to tail, The body undulation wave is
\begin{align}
    \alpha(\tau_b,i)=\Theta_{body} \text{cos}(\tau_b - 2\pi\frac{\xi^b}{n}(i-1)),
    \label{eq:latBodyUndu}
\end{align}

\noindent where $\alpha(\tau_b,i)$ is the angle of $i$-th body joint at phase $\tau_b$, $\xi^b$ denotes the number of spatial waves on body. For simplicity, we assume that the spatial frequency of the body undulation wave and the contact pattern wave are the same, i.e. $\xi^b = \xi$. 
In this way, gaits of multi-legged locomotors by superposition of a body wave and a leg wave can be described as the phase of contact, $\phi_c$, and the phase of lateral body undulation $\tau_b$. As discussed in~\cite{chong2022general}, the optimal body-leg coordination (optimal phasing of body undulation to assist leg retraction) is $\phi_c=\tau_b-(\xi/N+1/2)\pi$. 

\subsection*{Dynamic model}

As discussed in Eq. 2, the ground reaction force acting on $i$-th module  (a pair of legs and a body connection unit) is given by:

\begin{align}
    f_y^{i}(\tau,v) & = -\mu N \sin{\big(\tan^{-1}{(\frac{v_y(\tau,v)}{v_x(\tau)})}\big)} 
\end{align}

Thus, the total force acting on the $n$-link robot is:

\begin{align}
    f_y^{all}(\tau,v) & = -\mu N \sum_{i=0}^{n-1} \sin{\big(\tan^{-1}{(\frac{v_y(\tau+2\pi\frac{\xi}{n}i,v)}{v_x(\tau+2\pi\frac{\xi}{n}i)})}\big)} 
\end{align}

Now we replace $\tau$ as $2\pi f t$ (t is time), $N$ as $mg/n$ (m is the mass of the robot), $f_y^{all}$ as $-m\dot{v}$, we have:

\begin{align}
    \dot{v}(t,v) & = \frac{\mu g}{n} \sum_{i=0}^{n-1} \sin{\big(\tan^{-1}{(\frac{v_y(2\pi f t+2\pi\frac{\xi}{n}i,v)}{v_x(2\pi f t+2\pi\frac{\xi}{n}i)})}\big)}, 
\end{align}

\noindent which we can solve numerically to get the dynamic simulation.

\subsection*{Lego field built up}

We built up an obstacle rich environment using Lego blocks. First, we divided an 150cm $\times$ 90cm area of a wooden sheet into 1.5cm $\times$ 1.5cm squares (100 blocks along the length, 60 blocks along the width). The distribution of those Lego bricks\footnote{Included in LEGO Classic Large Creative Brick Box: https://www.lego.com/en-us/product/lego-large-creative-brick-box-10698} were generated by MATLAB \textbf{\textit{rand}()} function. Specifically, we used \textbf{\textit{rand}()} to create a $100\times 60$ matrix with uniformly distributed random variable. We marked the row and column information for the entries with highest 120 values. We placed the Lego bricks to the designated position. Finally, we hot glued Lego bricks on planned positions. 

\bibliography{sorsamp}

\end{document}